\documentclass[12pt,preprint]{aastex}

\usepackage{graphicx}

\usepackage{natbib}

\newcommand{\be}{\begin{equation}}
\newcommand{\ee}{\end{equation}}
\newcommand{\nn}{\mbox{} \nonumber \\ \mbox{} }
\newcommand{\ba}{\begin{eqnarray}}
\newcommand{\ea}{\end{eqnarray}}
\newcommand{\om}{\omega}
\newcommand{\Alfven}{Alfv\'{e}n }

\newcommand\eg{\textit{e.g.}}

\newcommand{\Bf}{{magnetic field}}
\newcommand{\Bfs}{{magnetic fields}}

\newcommand{\ms}{magnetosphere}
\newcommand{\LC}{light cylinder}

\newcommand{\Sc}{Schwarzschild}
\newcommand{\BH}{black hole}

\newcommand{\cfg}{centrifugal}

\begin{document}

\title{Magnetocentrifugal launching of jets from   disks around Kerr black holes }
\author{Maxim Lyutikov}
\affil{
Department of Physics, Purdue University, 
 525 Northwestern Avenue,
West Lafayette, IN
47907-2036 }

\begin{abstract}
Strong magnetic fields  modify particle motion in the curved space-time of  spinning black holes and change the stability conditions of circular orbits.
We study conditions for  magnetocentrifugal jet  launching from   accretion disks around black holes, whereby  large scale black hole  lines anchored in the disk may fling  tenuous  coronal gas outward.
For a Schwarzschild black hole, magnetocentrifugal launching requires that the  poloidal component of magnetic fields  makes an angle less than $60^\circ$ to the outward direction at the disk surface, similar  to the  Newtonian case. For prograde rotating disks  around Kerr black holes, this angle increases and becomes  $90^\circ$ for  footpoints anchored to the disk near the horizon of a critically spinning $a=M$ black hole.  Thus, a disk around a critically spinning black hole  may centrifugally launch a jet  even along the rotation axis. 
 \end{abstract}

\maketitle

\section{Introduction}
\label{intro}

The production of relativistic  jets in (micro)quasars, Active Galactic Nuclei and Gamma Ray Bursts involves the conversion of  rotational energy of the accreting matter and/or of the spinning \BH\ into   plasma bulk motion \citep{BlandfordZnajek,BlandfordPayne,PunslyCoroniti}.
This energy conversion is mediated by  the \Bf,  which can both accelerate plasma by magnetic gradient forces and collimate it by hoop stresses,  \cite[see][for review]{Meier01}.
 It is not clear whether  jet is powered by the accretion disk or by the spinning \BH\ (more precisely, whether the footpoints of \Bf\ lines that eventually form the jet should pass throughout the disk, ergosphere or horizon of the \BH). Numerical simulations confirm, at least partially, that all these  mechanisms can be operational \citep{Meier01,Koide,Komissarov05,Tchekhovskoy08}.
 
 There are three  related issues in the production of jets: launching, acceleration and collimation. In this paper we address the  conditions for jet launching from a thin accretion disk rotating in the equatorial plane of a Kerr \BH.  We assume that the disk is permeated by a large scale \Bf, so that magnetic field lines are anchored in the disk and are dragged both by plasma rotation and the rotation of space time around the \BH. This assumes that the disk conductivity is very high,  and that inside the disk the \Bf\ is subdominant and does not affect plasma motion in the disk. In addition, we assume that the  disk is surrounded by a  tenuous coronal plasma and the \Bf\ is  sufficiently  strong to affect particle motion in the corona.  The dynamical effects of  a  large scale \Bf\ on  single
particle motion are often approximated as a  guiding wire with a particle playing the role of a  bead. This  simple approximation restricts particle motion across the field, 
 neglecting various cross-field drifts. 
 
  We study the dynamical behavior of  coronal particles, located close to the disk, which are  {\it locally}   restricted to move exclusively along a given \Bf\ line. In such a case,  under certain conditions discussed below, a coronal particle executing a circular motion around a  \BH\ is in a state of unstable equilibrium, even though without  a \Bf\ it may be in a stable equilibrium.  The condition of unstable equilibrium, we assume, constitutes a condition for jet launching from an accretion disk.  This  may be the most relevant condition  for the generation 
of jets \citep{BlandfordPayne,OgilvieLivio,Spruit08}.


\section{Bead on a wire rotating in Kerr space}

\subsection{Rotating Kerr metric}

\begin{figure}[h]
\includegraphics[width=.99\linewidth]{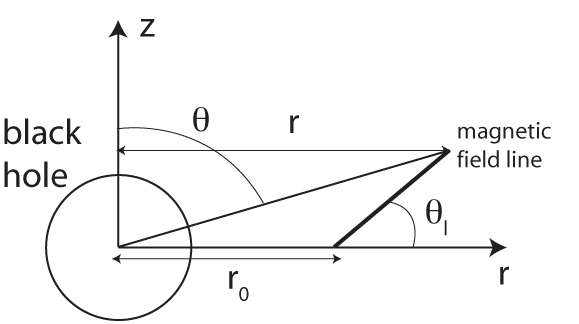}
\caption{Geometry of the model. A magnetic field line is anchored in the disk at the equatorial plane of a Kerr \BH\ at radius $r_0$ and makes an angle $\theta_l$ (in  Boyer-Lindquist coordinates) with the radial direction.}
  \label{footpoint}
\end{figure}

Assume that a footpoint of a \Bf\  line is anchored in accreting matter and is rotating with angular velocity $\om$ at $r=r_0$. Next,  neglect the toroidal component of the \Bf\ (see Appendix \ref{toroidal}) and assume that a field line makes an angle $\theta_l$ with the radial direction, Fig. \ref{footpoint}

As a mathematical problem, consider the motion of a  bead on a wire sticking out of the  equatorial plane of a Kerr \BH\ at an angle $\theta_l$ with the radial direction and confined to move in a given meridional plane.
In Boyer-Lindquist coordinates \citep[\eg][]{MTW}, make  a coordinate transformation  to the frame rotating with the foot-point, $\phi \rightarrow \phi' - \om t$. Restricting particle motion along the wire,
$\tan \theta= r/((r-r_0) \tan \theta_l)$, $d\phi=0$ (see Fig. \ref{footpoint})
implies 
\be
d\theta= \cot \theta_l  { r_0 dr \over (r-r_0)^2+r^2  \cot ^2\theta_l}.
\ee
In the coordinate frame rotating with the wire
 the non-vanishing components of the metric tensor are
 \ba &&
 g_{00}= - 1 + {1 \over \Sigma} \left(  {2 M r} + { 4a M r^3 \om \cot ^2 \theta_l \over\Delta_2} + \left( (a^2+r^2)^2 - { a^2 r^2 \cot ^2 \theta_l \over \Delta_2}\right) { r^2 \om^2 \cot^2 \theta_l\over \Delta_2}
 \right) 
  \nn &&
 g_{rr}= \left( {1 \over \Delta} + {r_0^2 \cot ^2 \theta_l \over \Delta_2^2} \right) \Sigma
 \nn &&
 \Delta = a^2 +r^2- 2M r
 \nn &&
 \Sigma = r^2+ a^2 \cos^2 \theta
 \nn &&
 \Delta_2= (r-r_0)^2 + r^2 \cot ^2 \theta_l
 \label{Kerr}
 \ea
where $a$ is the angular momentum per unit \BH\ mass and $M$ is the  mass of the  \BH. (We set $G=c=1$, use $(-1,1,1,1)$ sign convention and assume that the mass of a test particle is unity.)
 We assume that $\om >0$, while the parameter  $a$ can be in the range $-M <a<M$ ($a>0$ corresponds to prograde rotation of the wire).
Below we refer to the metric (\ref{Kerr}) as a  rotating Kerr metric. 

Condition $g_{00}=0$ defines two  light cylinders. Let us consider the location of the {\LC}s for a particular case of radial \Bf\ $\theta_l=0$, $\theta =\pi/2$. 
In this case 
\ba &&
g_{00} = -1 -(a^2 +r^2)\om^2 + {2 M (1- a \om)^2 \over r}
\nn &&
g_{rr}= { r^2  \over \Delta}.
\label{met}
\ea
 For $\om=0$, the inner \LC\ coincides with the ergosphere,  $r=2M$. For small $\om<< 1/M$, the outer \LC\ is approximately at $r\sim 1/\om$. The location of  the  {\LC}s are given by 
\ba &&
r_{\rm in} = \sqrt{{1\over \om^2}-a^2} \left( \sin  \left({1\over 3} {\rm arccos} \left( -
 {3 \sqrt{3} (1- a\om)^2 M \om \over (1-a^2 \om^2)^{3/2}  } \right) \right)-
  \right.  \nn && \left. 
 {1\over \sqrt{3}}    \cos  \left({1\over 3} {\rm arccos}\left(  -  {3 \sqrt{3} (1- a\om)^2 M \om \over (1-a^2 \om^2)^{3/2} } \right)\right)
\right)
\nn &&
r_{\rm out} = {2\over \sqrt{3}} \sqrt{{1\over \om^2}-a^2} \cos \left({1\over 3} {\rm arccos} \left( -  {3 \sqrt{3} (1- a\om)^2 M \om \over (1-a^2 \om^2)^{3/2} } \right) \right)
\ea
  Since the determinant of the metric tensor is smaller than zero beyond the {\LC}s, the approximation of a rigidly rotating wire is inapplicable in those regions.
 
 Condition $\partial_r g_{00}=0$ defines circular orbits
 \be
 r_0^3 \om^2 = M (1-a \om)^2
 \label{rc}
 \ee
 This is Kepler's law in a  Kerr metric.
 
\begin{figure}[h!]
\includegraphics[width=1\linewidth]{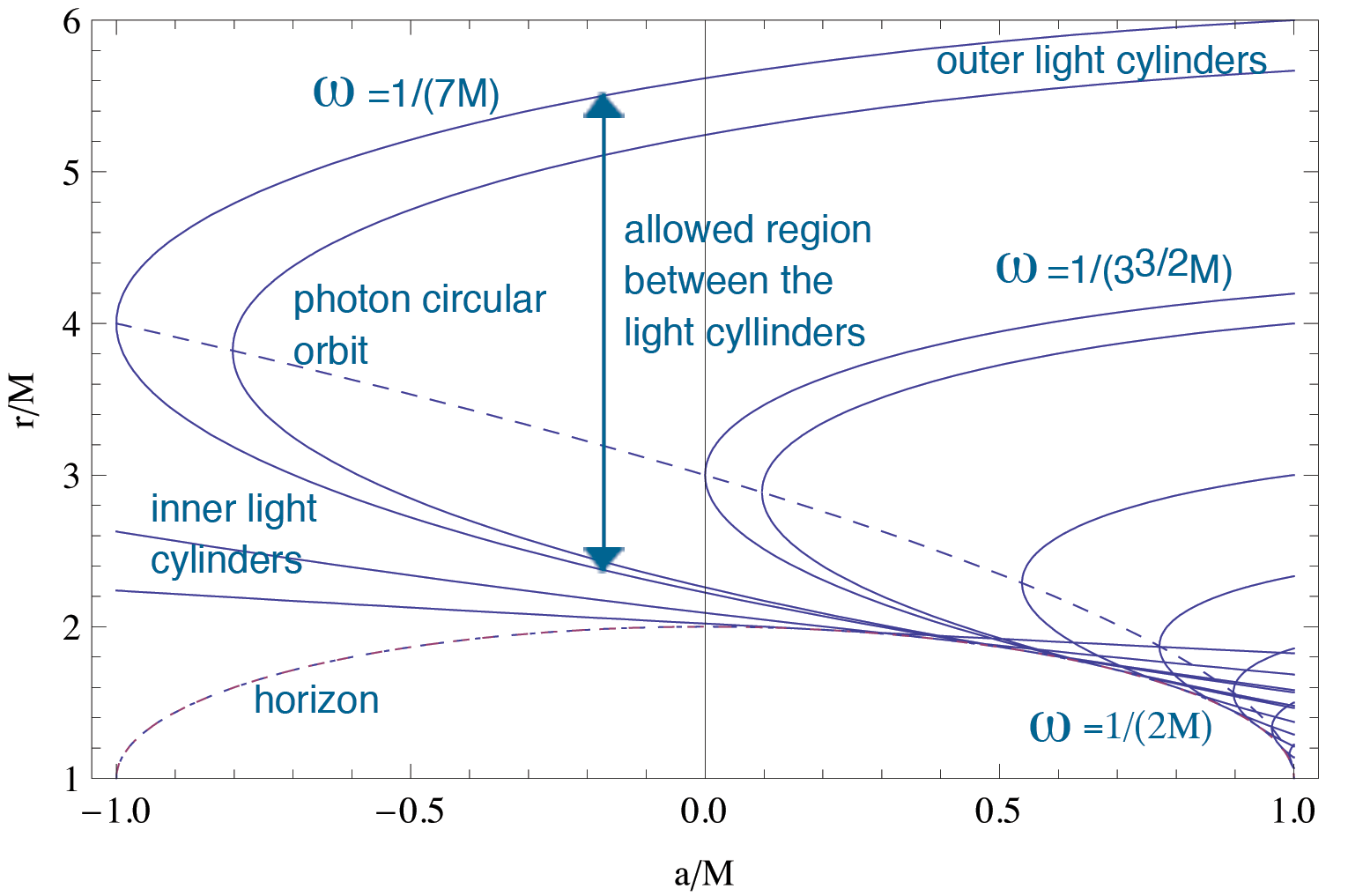}
\caption{The locations of {\LC}s for different $\om$ as function of $a$.  The light cylinders always lie outside the horizon (dotted curve). For a given $\om$, the inner and outer {\LC}s merge at  the photon circular orbit (dashed line).
 The physically meaningful region lies in between the  {\LC}s,   to the right of the {\LC}s curve. For $\om \rightarrow 0$, the inner \LC\  coincides with the ergosphere $r=2M$.  As $\om$ increases, the outer \LC\ moves to smaller $r$; the radial location of the  inner \LC\  is a complicated function of $\om$.  For $\om > 1/(7M)$ there is a region for sufficiently small $a$, for which the transformation to the rotating frame is unphysical.
For high $\om \rightarrow 1/(2M)$ and high $a > M/\sqrt{2}$, the inner \LC\ moves inside the ergosphere of a Kerr \BH.  
 For $\om > 1/(2M)$,   the transformation to rotating frame is  unphysical everywhere.}
  \label{r(t)}
\end{figure}

 Transformation to the rotating Kerr metric is physical only  for angular velocities smaller than $\om_{ph}$,  angular velocity of a photon orbit, defined by the conditions of circular rotation with the speed of light $g_{00}=0, \, \partial_r g_{00}=0$.  This gives $-4 a^2 M + r (-3 M + r)^2=0$ \citep{Bardeen72}. 
For a given $a$ and $r$, the transformation to the rotating Kerr metric   becomes meaningless for $\om$ higher than the angular velocity of a photon circular orbit, 
\be
\om_{\rm ph } = { 1 \over \mp | a| + 6 M \cos({1\over 3} {\rm arccos} (\mp |a|/M)}
\label{k2}
\ee
The upper sign corresponds to prograde rotation. Particular values are $a=0,\, r_{\rm ph}=3M,\, \om _{\rm ph}=1/(3\sqrt{3} M)$ for \Sc\ \BH, 
$a=M,\, r_{\rm ph}=M,\,\om_{\rm ph}=1/(2M)$ for prograde and $a=-M,\, r_{\rm ph}=4M,\,\om_{\rm ph}=1/(7M)$ for retrograde photon orbits.
For $1/(7M) < \om < 1/(2M) $ this requires  sufficiently high $a$, satisfying  $\om < 1/\left( a +2 \sqrt{2} M \left(1+\cos \left( {2\over 3}  {\rm arccos} (-{a\over M} ) \right)\right)\right)^{3/2}$ (for $a=0$ this requires $\om  < 1/(3 \sqrt{3})$). For $\om \rightarrow 1/ (2 M)$ both {\LC}s merge on the \BH\ horizon at  $a=M=r$.  For higher $\om $, the  transformation to the  rotating frame becomes  meaningless everywhere (Fig. \ref{r(t)}).

 \subsection{Motion of  particle in rotating Kerr metric.}
Since  metric (\ref{met}) is time independent, the Hamilton - Jacobi equation,
\be
g^{\mu \nu} \partial_\mu S \partial_\nu S= { g^{00}} (\partial_t S)^2 +{ g^{rr}} (\partial_r S)^2 =-1
\ee
has  a solution in the form $S=-E_0 t + S_r(r)$ with
\be
S_r =\int dr \sqrt{- (E_0^2 + g_{00}){ g_{rr} \over g_{00}}}
\ee
Differentiating with respect to $E_0$ gives the equation of motion
\be
(\partial_t r)^2 = -(1 + g_{00}/E_0^2)  {g_{00} \over g_{rr}}   
\ee
Transforming to proper time $dt/d\tau =E_0/(-g_{00})$ gives
\be
(\partial_\tau r)^2 = - {E_0^2 +g_{00} \over g_{00} g_{rr}} \equiv V
\label{rtau}
\ee
 This defines the effective potential $V$. 
 The extremum of the potential $\partial_r V=0$ defines the angular velocity of a  circular orbit. Direct calculations give, naturally, condition (\ref{rc}). 
 

\subsubsection{Reversal of \cfg\ acceleration at photon circular orbit}

The above relation simply in  case of  a radial \Bf, $\theta_l=0$. In this limit 
\ba &&
V= -  {\Delta \over r^2} \left( 1- { E_0^2 \over 1 - {2 M \over r } (1- a \om)^2- \om^2 (r^2+a^2) }\right)=0\,  \, \mbox{(for circular orbits)}
\nn &&
\left. {\partial^2 r \over \partial \tau^2} \right|_{E_0 = \sqrt{-g_{00}} }= {1\over 2} \partial_ r V = -{ M r -a^2 \over r^2} +  \left({a+ (r^2+a^2) \om  \over r^2}  \right)   
{ a (M-r) + \om \left(r^2 (r-3M) + a^2 (r+M) \right) \over r^2}.
 \label{rtau2}
\ea
For a  \Sc\ \BH, $a=0 $, Eq. (\ref{rtau2}) shows the reversal of centrifugal acceleration  at the photon circular orbit  $r= 3M$ \citep{AbramowiczPrasanna}:
 \be 
{\partial^2 r \over \partial \tau^2} = 
 - { M \over r^2}+(r-3M) \om^2
 \label{a}
\ee
For a Kerr \BH,  no clear separation can be made between the effects of the wire rotation and the rotation of space-time, so that  the notion of a centrifugal force becomes ill-defined   \citep[see][ for discussion]{AbramowiczPrasanna,deFelice91,AbramowiczLasota97,IyerPrasanna}.

\section{Stability of circular orbits in \Bf}
\label{Stability}

Circular orbits are defined by $ \partial_\tau r=0$ ($V=0$, $ E_0=\sqrt{-g_{00}}$) and  $ \partial^2_\tau r=0$ ($V_r'=0$, condition (\ref{rc})).
  The stability of an orbit depends on 
the second derivative of the potential $V$, Eq. (\ref{rtau}). 
\be
\kappa^2= - \left. {1\over 2} V^{\prime \prime}_{rr} \right|_{E_0=\sqrt{-g_{00}}, r=r_0} = 
{\Delta \left(3 a^2 - 4 a \sqrt{ M r}  +r^2 (1-3 \cot^2 \theta_l) \right)\over (r^4(r-3M)+ 2 a \sqrt{M} r^{7/2} ) ( \Delta + r^2 \cot ^2\theta)} M,
\label{kappa}
\ee
where we replaced $r_0$ by $r$. Stability requires the  epicyclic frequency to be real, $\kappa ^2 >0$. We identify parameters for which $\kappa ^2<0 $ as a necessary condition for jet launching.

The sign of $\kappa ^2 $ in  Eq. (\ref{kappa}) determines whether a particle confined to move along the \Bf\ anchored at radius $r$ to the disk is in a state of stable equilibrium ($\kappa ^2 >0$).
In Eq. (\ref{kappa}),  the denominator changes sign at the photon circular orbit $ r^4(r-3M)+ 2 a \sqrt{M} r^{7/2} =0$, Eq. (\ref{k2}). The physically realizable case corresponds to radii larger than the radius of the photon circular orbit. The numerator in Eq. (\ref{kappa})  changes sign at 
\be
\cot \theta_l =\sqrt{ {1\over 3} - {  4 a \sqrt{M} \over 3 r^{3/2}} + {a^2 \over r^2} }
\label{thetal}
\ee
When the \Bf\ lines make an angle with the radial direction less than the one given by Eq. (\ref{thetal}), a particle tied to the field line in the equatorial plane is in a state of unstable equilibrium and will be flung away from the disk.  

An analysis of Eq. (\ref{thetal}) reveals, first, that the  \Bf\ cannot stabilize particle motion for an  arbitrary (especially  radial) direction of the  \Bf\ (angle $\theta_l$ is never zero). In  the \Sc\ metric, $\cot \theta_l =\sqrt{ 1/ 3}$, $\theta _ l = \pi/3$, which is exactly the same result as in Newtonian mechanics \citep{BlandfordPayne}.
For  prograde rotation,  $a> 0$, the  instability region is larger, $\theta _ l > \pi/3$, so that the flow is more unstable in the Kerr metric if compared with the  Newtonian and  \Sc\ cases. Or, equivalently, a flow my be launched over a wider  range of angles.

Physically meaningful cases correspond to footpoint motion along stable circular orbits in  a Kerr space-time.  For prograde rotation  around a critically spinning Kerr \BH,  the outer and inner {\LC}s,  photon circular orbit and    horizon all coincide at $r=M$ in Boyer-Lindquist coordinates; in proper radial coordinates  they remain distinct \citep{Bardeen72}. According to Eq. (\ref{thetal}),
in the limit  $a\rightarrow M$, the angle  $\theta _ l \rightarrow \pi/2$, so that the motion along all field lines, including those aligned with the axis of rotation, becomes unstable. 
Thus, in the case of  a critically spinning Kerr \BH,  a jet may be launched by magnetocentrifugal mechanism {\it along} the rotation axis of a \BH.

For any $a\neq M$, the  condition $\theta_l=\pi/2$ is satisfied only inside the horizon, so that for $a\neq M$  there is always a set of \Bf\ lines around the direction of hole's rotation where a  particle is in a state of stable equilibrium. A jet won't be launched along those field lines.

  \begin{figure}[h!]
\includegraphics[width=.99\linewidth]{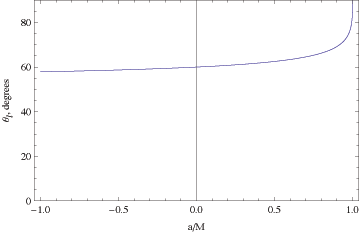}
\caption{The maximum unstable angle $\theta_l$ of the  \Bf\ inclination to the plane of the disk evaluated at the innermost  stable orbit . For angles less that $\theta_l$, a particle on a circular orbit is in  an unstable equilibrium and will be flung away by centrifugal forces. For a  \Sc\ \BH, $a=0$, $\theta_l=60 ^\circ$, similar to the Newtonian case. For $a \rightarrow M$,  a particle can be accelerated along the rotation axis.  For counter rotating disk around critically spinning \BH\ the critical angle decreases to cot$\theta_l =4 \sqrt{2}/9,\,\theta_l= 57.85^\circ$}
  \label{rms}
\end{figure}
\begin{figure}[h!]
\includegraphics[width=.99\linewidth]{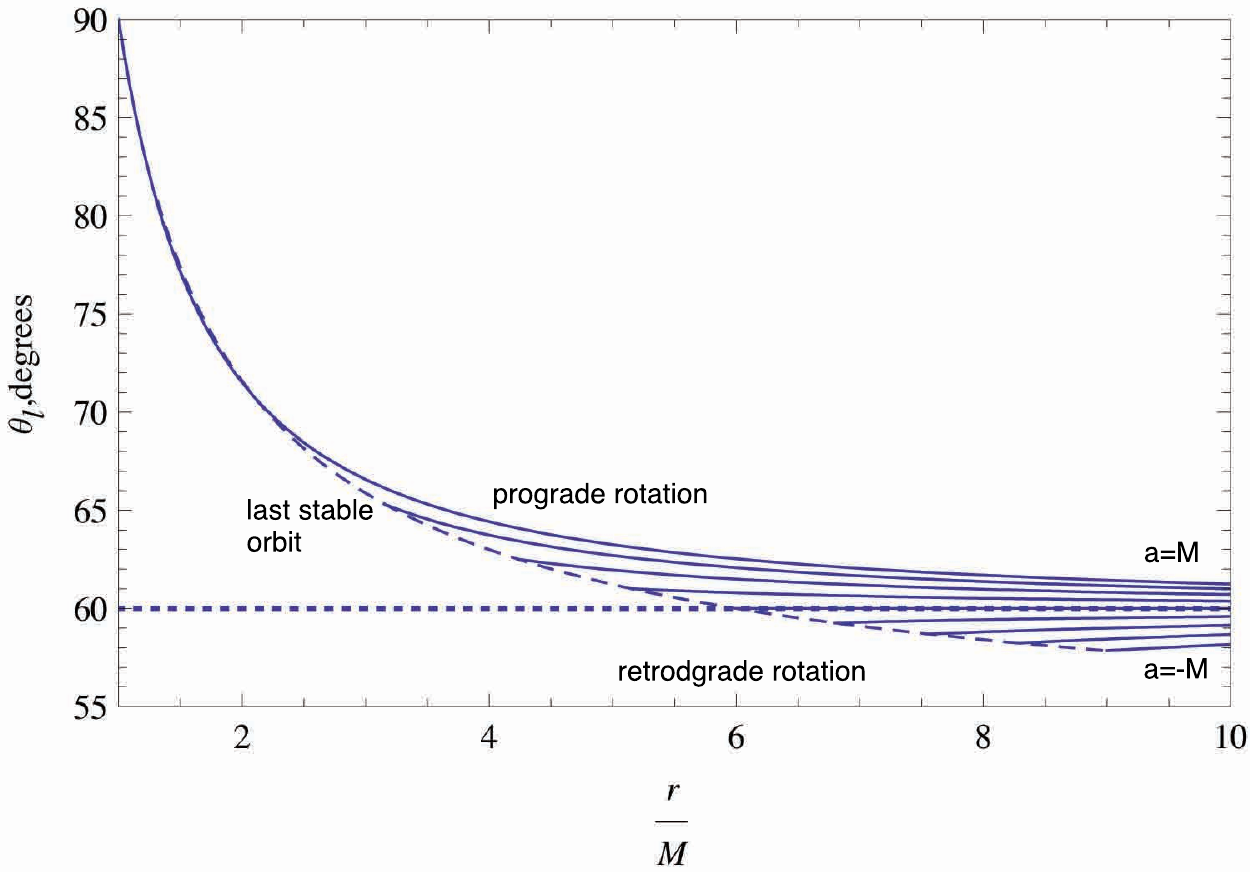}
\caption{Values of the  maximum unstable angle $\theta_l$ of the  \Bf\ inclination to the plane of the disk as function of radius for various values of \BH\ spin $a$. 
Dashed line is the values of $\theta_l$ at the innermost  stable orbit, see Fig. \ref{rms}. Dotted line corresponds to \Sc\ \BH, $a=0$,  $\theta_l=60^\circ$}
  \label{thetalofr}
\end{figure}

 At the innermost  stable orbit \citep{Bardeen72}, the  angle $\theta_l$ is, generally, close to $60^\circ$, except in a narrow  region of parameter $a\sim M$ for prograde rotation, where $\theta_l$ approaches $90^\circ$, see  Fig. \ref{rms}. For  a retrograde disk  rotating  around a critically spinning  \BH, the maximum launching angle decreases  to $\cot \theta_l =4 \sqrt{2}/9,\, \theta_l=57.8 ^\circ$ at the last retrograde  stable orbit corresponding to $r=9M$.
 In Fig. \ref{thetalofr} we plot the values of the  maximum unstable angle $\theta_l$ of the  \Bf\ inclination to the plane of the disk as function of radius various values of \BH\ spin $a$.

\section{Force-free magnetospheres}


Conditions for stability of particle motion at the base of the jet  derived in \S \ref{Stability} should be applied to the magnetic structures of \BH\ {\ms}s to determine if a particular  \Bf\ solution is consistent with the jet launching conditions at the disk surface. Analytical models of \BH\ {\ms}s are usually limited to the   force-free approximation, when matter inertia and pressure forces are neglected. The equation governing  force-free magnetic configurations in the  Kerr metric was derived by  \cite{BlandfordZnajek} (see also \cite{ThornMembrane}).
No  analytical solution for the magnetic field structure of fast spinning {\BH}s, $ a \sim M$, is known.  (The vacuum Wald solution \citep{Wald} has large parallel electric fields; this limits its astrophysical applicability. In addition, the Wald field is orthogonal to the equatorial plane.)

 Analytical solutions 
of  force-free structures for \Sc\  and slowly spinning Kerr \BH\ {\ms}s ($a \rightarrow 0$) are generally limited to the case of zero poloidal current.\footnote{One may also find spheromak-type solutions for linear dependence of the poloidal current on the poloidal magnetic flux, but they do not seem to correspond to any physically interesting case.} In this case ($a=0$) the governing Grad-Shafranov equation   \citep{Shafranov} for the poloidal flux function $P$ becomes  \citep{BlandfordZnajek,ThornMembrane}
\be
r^2 \partial_r \left( (1-2M/r)  \partial_r P \right) +(1-\mu^2)  \partial_\mu ^2 P=0
\label{Grad}
\ee
where $\mu = \cos \theta$. Magnetic fields measured by  a locally stationary observer are then given by \citep{ThornMembrane}
\ba &&
B_{\hat{r}} = {\partial _\theta P \over r^2 \sin \theta}
\nn &&
B_{\hat{\theta}}= \sqrt{1-2M/r}  {\partial _r P \over r \sin \theta}
\label{B}
\ea
(Note, that a a form of the flux function $r(\theta)$  as measured by a set of locally stationary observer is given by $dr/d\theta=B_{\hat{\theta}}/B_{\hat{r}} = 
 \sqrt{1-2M/r} ( r \partial _r P / \partial _\theta P)$, not just $P={\rm constant} $ as in the flat space). 

There is a number of known solutions of Eq. (\ref{Grad}). Those connecting to the accretion disk (as opposed  exclusively to the \BH)  are (i) a  \Sc\ \BH\ in the constant \Bf\ $P  \propto r^2 \sin ^2 \theta$; (ii) the non-separable parabolic  solution of \cite{BlandfordZnajek} $P\propto  (r-2M)(1-\mu) - 2M (1+\mu) \ln (1+\mu)$.;
(iii) the separable Schwarzschild paraboloid $P\propto \left(r+ 2M \ln (r-2M) \right) \left(1-\mu\right)$ \citep[and corresponding  higher order multipoles][]{Ghosh00}.
These are all linear solutions, so that any combination is a possible solution as well. 

\begin{figure}[h!]
\includegraphics[width=.99\linewidth]{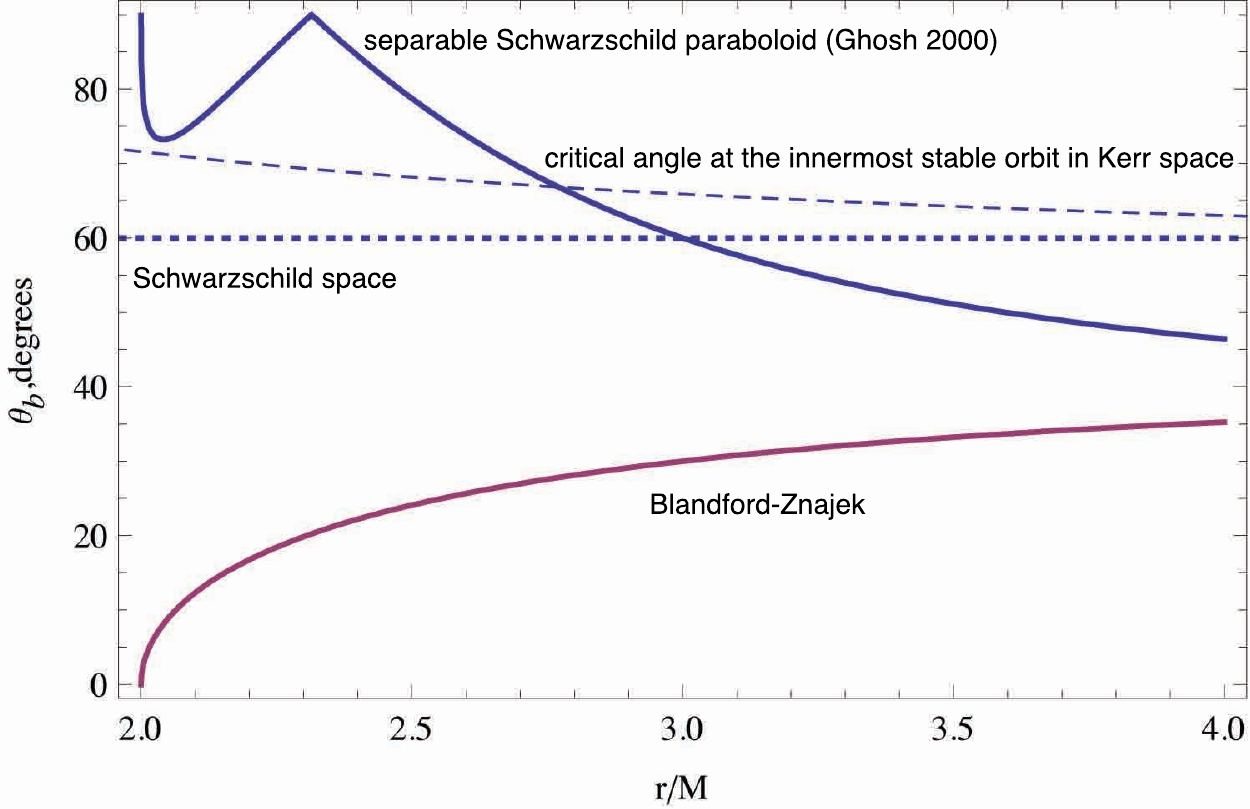}
\caption{Values of the angle $\theta_b$ between  \Bf\  and the equatorial plane  for   force-free {\ms}s of \Sc\ \BH: \cite{BlandfordZnajek} and separable Schwarzschild paraboloid \citep{Ghosh00}. These solutions, and a solution for a \BH\ in constant \Bf, are linear solutions, so that any combination is a possible solution as well.  A given solution can centrifugally launch a jet if the angle $\theta_b$ is smaller than the critical angle at the innermost stable orbit. }
  \label{Bfieldangle}
\end{figure}

The angle $\theta_b$ that \Bf\ lines make with the equatorial plane is 
\be
\tan \theta_b = - {B_{\hat{\theta}} \over B_{\hat{r}}} =   \sqrt{1-2M/r} \left. { r \partial _r P \over \partial _\theta P}\right|_{\theta = \pi/2}
\ee
The constant \Bf\ solution (i) has a  \Bf\ orthogonal to the disk at all points.
 For the  parabolic solution of \cite{BlandfordZnajek} (ii),  the angle that the  \Bf\ lines make with the equatorial plane decreases as the footpoint approaches the horizon, $\tan \theta_b  = \sqrt{1-2M/r}$; for large radii, $r\gg M$, field lines stick out  at $45^\circ$ to the plane of the disk. Thus, for  the parabolic solution of  \cite{BlandfordZnajek}, the particles confined to move along  field lines are always in an unstable equilibrium, so that the  jet can be launched from any point on the disk. For the separable Schwarzschild paraboloid  solution (iii),  the angle that \Bf\ lines make with the equatorial plane is $\tan \theta_l =\left(  \sqrt{1-2M/r} (1+(2M/r) \ln (r-2M))\right)^{-1}$.  It reaches  $\theta_l=\pi/2$ somewhere close to the horizon (at the root of $1+(2M/r) \ln (r-2M))=0$) and then again at the horizon $r=2M$, see Fig. (\ref{Bfieldangle}).
Thus, the force-free \Bf\ structures of \Sc\ {\BH}s allow  for a variety of field configurations with different inclinations of \Bf\ lines at the disk surface.  We may only speculate that Kerr holes have a  \Bf\ structure with various \Bf\ inclinations at the disk surface as well. 

\section{Conclusion}

In this paper we discussed the conditions for the  centrifugal  launching of jets from a thin accretion disk around spinning {\BH}s. 
Typically, it is believed that ``for acceleration by the centrifugal mechanism to be effective, the field  lines have 
to be inclined outward'' \citep{Spruit08}. This leads to a collimation problem: a flow should first accelerate by magnetocentrifugal forces  outwards in cylindrical radius and later, beyond the \Alfven surface, it needs to be collimated along the axial direction by hoop stresses of bent back \Bf\ lines \citep{BlandfordPayne}. For relativistic motion, such collimation becomes very ineffective \citep{Eichler93}.

We found that \Sc\ {\BH}s conform to these expectations: conditions for jet launching (defined as an instability of a particle on a  circular orbit confined to move along  a \Bf\ line) remain the same as for Newtonian disks: a field should be inclined to the disk surface at an angle $\theta_l < 60 ^\circ$ (when viewed in Boyer-Lindquist coordinates)  at any radius in order to launch a jet. On the other hand, in the case of a critically spinning Kerr \BH, the  centrifugal acceleration along \Bf\ lines anchored  in the disk may occur virtually {\it along the rotation axis of the \BH}.  Qualitatively, this removes the need for a flow to expand sideways to get accelerated.

The present result has implications for the transport of angular momentum by large scale \Bfs. Since the particle orbits in the inner regions of accretion disks around spinning  black holes are unstable to large scale magnetic fields with a wide range of angles at  the disk surface,  this make the  transport angular momentum away from the disk more efficient than in the case of \Sc\ {\BH}s.

The centrifugal acceleration mechanism  (a fling)  isn't the only way to magnetically  launch a jet. Alternatively, a jet may be launched by stresses of rotating twisted \Bf (a spring),
\cite[see, \eg][]{Tchekhovskoy08}. In a ``spring''-type launching the disk or the  central \BH\ act as a Faraday disk, which launches an outflow with the  terminal speed equal, approximately, to the \Alfven velocity in the corona. If the coronal plasma is strongly magnetized, the terminal velocity may be relativistic. 

It is tempting to relate the possibility of   \cfg\ acceleration along the  rotation axis in critically spinning {\BH}s  to the results of numerical simulations of jet launching.
It is well established  that the efficiency of jet production is a strong function of the \BH\ spin \citep{Meier01}; for a rapidly spinning {\BH}  the efficiency of converting the accreting energy  into jet power reaches
 the radiation efficiency of thin disks \citep{DeVilliers03,DeVilliers05}. We speculate that this is because jet launching for critically spinning \BH\ 
  is more efficient due to effect discussed in this paper.

I would like to thank Eric Brawn and Jonathan McKinney for comments on the manuscript. 

\bibliographystyle{apj}
\bibliography{/Users/maximlyutikov/Home/Research/BibTex}

\appendix

\section{Effects of toroidal component of \Bf}
\label{toroidal}

It is expected that for relativistic flow with rotation rates nearing the speed of light, the toroidal component of the magnetic field is not negligible. In this Appendix we show that inclusion of the toroidal component does not change the stability conditions described above. First note, that in the  case of Keplerian (non-realtivistic) rotation, the azimuthal component of the wire direction does not change the stability condition, as can be trivially shown (both centrifugal  force and the force of gravitational attraction do not have $\phi$-component, and thus their projections on any oblique direction become equal for the same parameters, as in the case of no toroidal component of the wire).

 In Kerr metric, we can repeat the previous derivation allowing for finite displacement in the azimuthal direction, along a direction given by angle $\phi_l$ with respect to the radial direction. For a given redial displacement $dr$ the change in the azimuthal angle is
\be
d\phi = \cot \theta_l  { r_0 dr \over \sin\theta_l \left((r-r_0)^2+r^2  \cot ^2\theta_l\right) } \tan \phi_l
\ee
The non-zero components of the metric tensor are $g_{00}, \, g_{rr},\, g_{rt}$ (not given here explicitly).
It is important that  in the rotating Kerr coordinates, inclined with respect to radial direction, the $g_{00}$ component of the metric tensor remains the same, Eq. (\ref{Kerr}).

Solution of the Hamilton-Jacobi equations are now
\be
S=-E_0 t -  \int dr {E_0 g_{rt} \pm \sqrt{ -(E_0^2 +g_{00})(g_{00} g_{rr} - g_{rt}^2) }\over g_{00} } 
\ee
Equation of motion is
\be
\left( {dr \over dt} \right)^2 =\left( {g_{rt} \over g_{00}} \pm {E_0 \over g_{00} } \sqrt{ g_{00} g_{rr} - g_{rt}^2 \over E_0^2 +g_{00}} \right)^{-2}
\ee
Transformation to proper time 
\be
{d\tau \over dt} = -{  g_{00} \sqrt{ g_{00} g_{rr} - g_{rt}^2}\over  \sqrt{ -(E_0^2 +g_{00})} g_{rt} - E_0 \sqrt{ g_{00} g_{rr} - g_{rt}^2}} 
\ee
 gives the effective potential
\be
\left( {dr \over d\tau} \right)^2 = - {E_0^2 + g_{00} \over  g_{00} g_{rr} - g_{rt}^2}\equiv V
\ee
Conditions of circular orbit give  $V=0$ ($E_0= - \sqrt{-g_{00}}$), $V'_r \propto g_{00,r}=0$. Thus, similar to the case of radial wire, the epicyclic  frequency of an oblique wire  is proportional to $ g_{00,rr}$. It changes sign at the same vertical angle $\theta_l$  as in the case $\phi_l=0$. Eq.  (\ref{thetal}).

\section{Pseudo-potentials}

General relativistic effects are often modeled using  pseudo-Newtonian potentials. For \Sc\ {\BH}s such a potential was 
first proposed by 
  \cite{PaczynskyWiita}, $\Phi_{PW}=M/(r-2M)$. For  the  Paczy\'{n}ski-Wiita  potential  and centrifugal potential $(1/2) \om^2 r^2$, the angular velocity corresponding to circular orbits is $\om = \sqrt{M/(r_0 (r_0-2M)^2)} $ and the
critical angle  
\be
\tan ^2 \theta_{PW} = 3+{ 4M \over r- 2M} 
\ee
This value is close to the exact value in \Sc\ geometry. Even at the last stable orbit ($r=6M$, $\tan ^2 \theta_{PW}=4$) it differs from $60^\circ $ by only $3.43^\circ$ degrees.
Thus,  in the case of \Sc\ \BH\ the Paczy\'{n}ski-Wiita potential reproduces reasonably well the launching conditions at the disk surface.

In principle, general relativistic effects should also be taken into account in calculation of centrifugal acceleration. Using relation (\ref{a})  for the centrifugal force, the   pseudo-centrifugal potential becomes  $(1/2) \om^2 (r-3M)^2$; the angular velocity corresponding to circular orbits is then  $\om = \sqrt{M/((r_0-3M) (r_0-2M)^2} $ and the 
critical angle  
\be
\tan ^2 \theta_{PW} = 3+{ 4M \over r- 2M} +{ 3M \over r- 3M} 
\ee
At the last stable orbit (assuming it is at $r=6M$) $\tan ^2 \theta_{PW}=5$, again reasonably close to the exact value $\tan ^2 \theta=3$.

\end{document}